\documentclass[prfluids,aps,10pt]{revtex4-2}
\usepackage{graphicx}
\usepackage{epstopdf, epsfig}
\usepackage{bm}
\usepackage{amsmath}
\usepackage{amsfonts}
\usepackage{tikz-cd}
\usepackage{scalerel}

\usepackage{natbib}
\usepackage{hyperref}
\hypersetup{
    colorlinks = true,
    urlcolor   = blue,
    citecolor  = black,
}
\def\ba{\bm{a}}
\def\bx{\bm{x}}
\def\bu{\bm{u}}
\def\bk{\bm{k}}

\def\bphi{\bm{\varphi}}

\def\bXi{\bm{\Xi}}
\def\barbphi{{\bar{\bphi}}}
\def\barphi{{\bar{\varphi}}}
\def\barbu{{\bar{\bm{u}}}}
\def\baru{{\bar{u}}}
\def\e{\mathrm{e}}
\def\d{\mathrm{d}}

\def\lie{\mathcal{L}}
\def\id{\mathrm{id}}
\def\i{\mathrm{i}}
\def\eps{\varepsilon}
\def\beq{\begin{equation}}
\def\eeq{\end{equation}}
\renewcommand\Re{\mathrm{Re}\,}

\DeclareRobustCommand*\vp{{\stretchrel*{\dag}{X}}}

\def\bxi{\bm{\xi}}
\renewcommand{\phi}{\varphi}

\newcommand{\dt}[2]{\frac{\d#1}{\d#2}}

\newcommand\av[1]{\langle #1 \rangle}

\def\bcdot{\cdot}
\def\bnabla{\nabla}

\newcommand{\RomanNumeralCaps}[1]
\linenumbers

\begin{document}
\preprint{APS/123-QED}

\title[Volume-preserving Lagrangian averaging]{Volume-preserving Lagrangian averaging using polar factorization}

\author{Abhijeet Minz}
\author{Lois E. Baker}

\author{Jacques Vanneste}\email[Email address for correspondence: ]{J.Vanneste@ed.ac.uk}%
 
\affiliation{ 
School of Mathematics and Maxwell Institute for Mathematical Sciences, University of Edinburgh, Edinburgh EH9 3FD, UK
}%

\date{\today}

\begin{abstract}
The generalised Lagrangian mean (GLM) theory of Andrews \& McIntyre provides a powerful framework to study the interactions between waves and flows. A drawback of this theory is that the Lagrangian mean velocity is divergent even for incompressible fluids because the mean flow map, which sends the Lagrangian labels of fluid parcels to their mean positions, does not preserve volume. This results, for instance, in vortices shrinking under Lagrangian averaging.
 
We overcome this drawback by revising the definition of the mean flow map, choosing it  as the volume-preserving map closest to the ``bare'' GLM mean map. A standard result of optimal-transport theory then shows that the new mean map is the volume-preserving factor in the polar factorization of the GLM mean map. 
We develop and implement a numerical method for the computation of the corresponding Lagrangian mean fields from simulation data.
The implementation builds on recently developed algorithms for the on-the-fly computation of Lagrangian means using the exponential and Butterworth filters. 
 
We demonstrate the value of volume-preserving Lagrangian averaging in simulations of the two-dimensional incompressible and shallow-water models. We compare the Lagrangian-mean fields obtained with and without the volume-preservation constraint. 
\end{abstract}

\maketitle

\section{Introduction}

Averaging is a key tool for the mathematical and numerical modelling of fluid flows with multiple time scales. Standard averaging is Eulerian, that is, carried out at fixed spatial location. An alternative is Lagrangian averaging, carried out at fixed particle label, that is, along particle trajectories.
Lagrangian averaging has conceptual and practical advantages over Eulerian averaging. Conceptually, Lagrangian averaging leads to averaged dynamical equations that are often simpler and closer in structure to the original unaveraged equations than those obtained with Eulerian averaging \citep{bret71,soward1972kinematic,andrews1978exact,grim84,holm02a,holm02b,buhler1998non,salmon2013alternative,buhler2014waves}. Practically, Lagrangian averaging avoids the entanglement between fast and slow motion that results from their mutual advection. In particular, it eliminates the blurring of flow features that fast advection introduces in Eulerian averages. This enables a clean separation of the flow into slow and fast components each often associated with vortical and wave motion \cite{Shakespeare2017c,Shakespeare2021a,jonesUsingLagrangianFiltering2023,baker2024lagrangian,wang2026wave}.

The generalised Lagrangian mean (GLM) theory of \citet{andrews1978exact} provides a framework for the formulation of Lagrangian-averaged models. A key feature is the replacement of particle labels by Lagrangian mean positions as independent variables for the Lagrangian mean fields. The Lagrangian mean positions are usually defined via a straightforward component-wise average of the actual positions. While this definition has the advantage of simplicity, it introduces the so-called divergence effect \citep{mcintyre1988note,buhler2014waves}:  for an incompressible fluid, the Lagrangian mean velocity is divergent even though the actual velocity is divergence free. Equivalently, the map between mean and actual positions does not preserve volume. 

To illustrate the divergence effect, consider a two-dimensional incompressible fluid in rigid-body rotation. The trajectories $\bx(t) = (x(t),y(t))$ of particles are circles. It is intuitively clear that the component-wise average 
\beq
\bar {\bx}(t) = (\bar x(t),\bar y(t)) = T^{-1} \int_{t-T/2}^{t+T/2} (x(s),y(s)) \, \d s,
\label{tophat}
\eeq
where $T$ is the averaging time, defines a mean position $\bar{\bx}(t)$ with 
\beq
| \bar{\bx}(t)| < |\bx(t)|,
\eeq
see figure \ref{fig:circle}.
Thus time averaging contracts circular trajectories. A simple computation in appendix \ref{app:rigid}  confirms this conclusion and extends it to a broad class of time averages. 

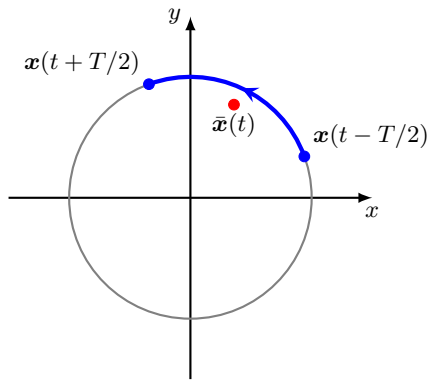
\begin{figure}
\begin{tikzpicture}[scale=0.8]

    \draw[-latex, thick] (-3, 0) -- (3, 0) node[below, text=black] {$x$};
    \draw[-latex, thick] (0, -3) -- (0, 3) node[left, text=black] {$y$};

    \draw[thick, gray] (0,0) circle (2);

    \draw[ultra thick, blue] (20:2) arc (20:110:2);
    \draw[ultra thick, blue,-stealth] (62:2) arc (62:65:2);

    \filldraw[blue] (20:2) circle (2.5pt) node[above right, text=black] {$\bx(t-T/2)$};
    \filldraw[blue] (110:2) circle (2.5pt) node[above left, text=black] {$\bx(t+T/2)$};
    \filldraw[red] (65:1.7) circle (2.5pt) node[below, text=black] {$\bar{\bx}(t)$};

\end{tikzpicture}
\caption{Time average of the position of a particle on a circular trajectory: the component-wise average \eqref{tophat} defines a mean position $\bar{\bx}(t)$ closer to the centre of rotation than the averaged trajectory $\bx(t)$.}
\label{fig:circle}
\end{figure}

Alternatives to the component-wise averaging of positions have been proposed to remedy the divergence effect and obtain volume-preserving versions of Lagrangian averaging \citep{soward1972kinematic,sowa-robe10,gilbert2018geometric,vanneste2022stokes}. 
These have been introduced for perturbative analyses of wave--mean flow process. In this context, they can be regarded as small modifications of standard GLM that trade the simplicity afforded by volume preservation (a constant mean density) against other complications (the perturbation fields acquire non-zero means).  

In this paper, we revisit volume-preserving Lagrangian averaging from a  different perspective. The new perspective is brought about by the recent development of efficient, PDE-based methods for the numerical computation of Lagrangian averages from simulation data \citep{kafiabad2022grid,kafiabad2023computing,baker2024lagrangian,minz2025efficient}. These methods, used primarily as diagnostic tools, rely on the standard GLM definition of the mean position and therefore suffer from the divergence effect. Here we overcome this drawback by developing a numerical method for the computation of a volume-preserving Lagrangian average. 

The method is based on the idea put forward by \citet{gilbert2018geometric} of replacing the mean flow map (sending particle labels to mean positions) of GLM by the closest volume-preserving map. Closest is here understood in a straightforward mean-square sense. With this definition, we can exploit a seminal result of optimal transport theory \citep{brenier1991polar} and characterize the new mean map as the volume-preserving factor in the polar factorization of the GLM map. 

We obtain a practical formulation, based on the velocity of the flow rather than the (usually exceedingly intricate) flow map, by considering the time derivative of the polar factorization. This leads to a system of equations relating the divergence-free Lagrangian mean velocity to the flow velocity (\S\ref{sec:formulation}). This system is similar to that obtained for GLM in that it couples the Lagrangian mean velocity with the map between mean and actual positions, but it also involves an additional scalar field that appears in the polar factorization. 

The construction of a volume-preserving mean map via 
polar factorization is not perturbative, that is, it does not  require the common assumption that the flow consists of a large mean component plus small-amplitude perturbations. Under this assumption,
we recover results of Ref.\ \citep{gilbert2018geometric} and clarify the differences between the mean flows corresponding to GLM, to  polar factorization, and to the alternative volume-preserving construction in Refs.\ \citep{sowa-robe10,vanneste2022stokes} (appendix \ref{sec:smallamp}). We also show how a superficially appealing solution to the divergence effect -- the projection of the GLM velocity onto a divergence-free velocity -- fails because it leads to mean trajectories that drift away from the particle trajectories.

We  demonstrate the capability of the proposed volume-preserving formulation for the computation of Lagrangian time averages from simulation data (\S\ref{sec:exponential_mean}). We choose the exponential filter as averaging operator. This is the simplest filter in a family of sum-of-exponential filters which make it possible to compute averages alongside the dynamical equations in a truly on-the-fly fashion, through the solution of additional evolution equations \citep{minz2025efficient}. The volume-preserving implementation retains much of the simplicity of the original GLM implementation in Ref.\ \citep{minz2025efficient}, requiring only a modest additional computational effort.

We implement the volume-preserving exponential filter for simulations of a two-dimensional incompressible fluid (\S\ref{sec:2d}) and of a rapidly rotating shallow water system (\S\ref{sec:SW}). 
We use the incompressible case, in which there are no fast waves to filter out, to contrast the way in which GLM and volume-preserving averaging alter slow vortical motion. 
The rotating shallow-water flow superposes to a turbulent vortical flow a fast Poincar\'e wave which the Lagrangian averaging aims to filter out. Although the shallow-water model is compressible, we find benefits in imposing volume-preservation of the mean map in the form of better wave filtering. We relate this finding to the approximate geostrophic balance of Lagrangian mean flows in the rapidly rotating regime \citep[e.g.][]{kafiabad2021wave}.

Applications of the central idea of this paper, namely the polar factorization of the GLM mean map, are not restricted to the exponential filter. We exemplify this with an implementation of the 2nd-order Butterworth filter, a more advanced sum-of-exponential filter (appendix \ref{app:butter}). We expect that the idea can also be adapted to more general filters used for  online or offline computations of volume-preserving Lagrangian means.

\section{Formulation} \label{sec:formulation}

\subsection{GLM}

The motion of a fluid is determined by the flow map  $\bphi(\ba,t)$ which gives the position at time $t$ of particles labeled by $\ba$ in a reference configuration. Lagrangian averaging starts by defining an analogous mean flow map $\barbphi(\ba,t)$. In GLM, it is simply the component-wise average of the flow map:
\beq
\barphi^i(\ba,t) = \av{\varphi^i(\ba,t)}
\label{meanmap0}
\eeq
for $ i=1,\,2,\,3$. Since we restrict out attention to Euclidean space, we can treat maps as vectors and rewrite \eqref{meanmap0} as
\beq
\barbphi(\ba,t) = \av{\bphi(\ba,t)}.
\label{meanmap}
\eeq

In what follows we consider two types of averages:   time averages of the form 
\beq
\av{f}(t) = \int_{-\infty}^\infty k(t-s) f(s) \, \d s,
\label{timeav}
\eeq
where the kernel satisfies
\beq
\int_{-\infty}^\infty k(t) \, \d t = 1,
\eeq
and ensemble averages of the form
\beq
\av{f}(t) = \int k(\varpi) f(t,\varpi) \, \d \varpi,
\eeq
where $\varpi$ is a ensemble parameter with probability density $k(\varpi)$ such that
\beq
\int k(\varpi) \, \d \varpi = 1.
\eeq
Ensemble averages assume an ensemble of flow maps, $\bphi(\ba,t,\varpi)$; we drop the  explicit dependence on $\varpi$ in what follows, but it should be kept in mind. 

The Lagrangian mean velocity, which we denote by $\barbu$ (without a superscript L, to keep the notation light), is the velocity associated with the mean flow map:
\beq
\partial_t \barbphi(\ba,t) = \barbu(\barbphi(\ba,t),t).
\eeq
With \eqref{meanmap}, it is also the (componentwise) Lagrangian mean of the velocity in the sense that
\beq
\bar \bu(\barbphi(\ba,t),t) = \av{\bu(\bphi(\ba,t),t)}.
\label{meanu}
\eeq

The flow map is decomposed as
\beq
\bphi = \bXi \circ \barbphi,
\eeq
where $\bXi$ is the perturbation map, which depends on $\varpi$ for ensemble averaging. With this decomposition and for ensemble average the definition \eqref{meanu} of the Lagrangian mean velocity can be rewritten as
\beq
\barbu(\bx,t) =  \av{\bu(\bXi(\bx,t),t)}
\label{meanu2}
\eeq
by substituting $\bx$ for the mean position,  $\bx = \barbphi(\ba,t)$. The same substitution in \eqref{meanmap} gives
\beq
\bx = \av{\bXi(\bx,t)}.
\eeq
With $\bXi(\bx,t)=\bx + \bxi(\bx,t)$, this becomes
\beq
\av{\bxi(\bx,t)} = 0.
\label{zeromean}
\eeq
Eqs.\ \eqref{meanu2} and \eqref{zeromean} are taken as defining equations for GLM by \citet{andrews1978exact}. They do not hold for time averages, however, because of the time dependence of the substitution  $\bx = \barbphi(\ba,t)$. As a result, 
\eqref{meanu2} and \eqref{zeromean} only hold approximately provided that there is a time-scale separation between mean and perturbation.  

\subsection{Volume preservation} \label{sec:vp}

Mass conservation implies that the density $\rho(\bx,t)$ of a fluid is related to a fixed density $\rho_0(\ba)$ in label space according to
\beq
\rho(\bx,t) \, \d \bx  = \rho_0(\ba) \, \d \ba,
\eeq
where $\bx = \bphi(\ba,t)$.  
Expressing the volume element $\d \bx$ in terms of $\d \ba$ then gives
\beq
\rho(\bx,t) = \left| \frac{\partial \bphi}{\partial \ba} \right|^{-1} \rho_0(\ba).
\label{density}
\eeq
Here $| \partial \bphi / \partial \ba|$ with $| \cdot |$ denoting determinant is the Jacobian of the flow map. The continuity equation
\beq
\partial_t \rho + \nabla \cdot (\rho \bu) = 0 
\label{continuity}
\eeq
follows by time differentiation and use of the Jacobi identity.

For a homogeneous incompressible fluid, the densities $\rho$ and $\rho_0$ are constants which can be taken to be $1$: $\rho=\rho_0=1$. 
The flow map $\bphi(\ba,t)$ preserves volume and its Jacobian is unity. Because this constraint is nonlinear, its average is non-trivial and the mean map does generally not preserve volume. Hence,
\beq
\left| \frac{\partial \bphi}{\partial \ba} \right| = 1
\quad \textrm{but} \quad 
\left| \frac{\partial \barbphi}{\partial \ba} \right| \not= 1.
\eeq
Correspondingly, while the velocity of an incompressible fluid is divergence-free, the Lagrangian mean velocity is divergent,
\beq
\bnabla \bcdot \bu = 0 \quad \textrm{but} \quad 
\bnabla \bcdot \barbu \not= 0. 
\eeq
This is the divergence effect in GLM. It necessitates the introduction of a density variable $\tilde \rho(\bx,t)$ with
\beq
\tilde \rho(\barbphi(\ba,t),t) = \left| \frac{\partial \barbphi(\ba,t)}{\partial \ba} \right|^{-1} 
\eeq
or, equivalently since $\rho^{-1} = | \partial \bphi / \partial \ba| = |\partial \bXi/\partial \bx| | \partial \barbphi / \partial \ba| = 1$,
\beq
\tilde \rho(\bx,t) = \left| \frac{\partial \bXi(\bx,t)}{\partial \bx} \right|,
\eeq
which satisfies the continuity equation
\beq
\partial_t \tilde \rho + \bnabla \bcdot ( \tilde \rho \barbu) = 0
\eeq
analogous to \eqref{continuity}.

It is possible to modify the definition of the mean flow map, hence \eqref{meanu2} and \eqref{zeromean}, to eliminate the divergence effect. We propose to replaces the ``bare'' GLM mean map \eqref{meanmap} by the closest flow map that preserves volume, that is, to replace $\barbphi$ by 
\beq
    \barbphi_\vp(\ba,t) = \underset{\bm{\psi}: \left| {\partial \bm{\psi}}/{\partial \ba} \right| = 1}{\text{argmin}} \int \| \bm\psi (\ba,t) -  \barbphi (\ba,t) \|^2 \, \d \ba,
\label{min}
\eeq
where the integration is over the reference configuration. Here and in what follows we use the subscript $\vp$ to indicate fields associated with the volume-preserving construction. In \eqref{min} we  measure closeness in terms of an $L_2$ norm; other choices are possible.

A definition equivalent to \eqref{min} for ensemble averaging is
\beq
\barbphi_\vp(\ba,t) = \underset{\bm{\psi}: \left| {\partial \bm{\psi}}/{\partial \ba} \right| = 1}{\text{argmin}} \av{ \int \| \bm\psi (\ba,t) -  \bphi (\ba,t) \|^2 \, \d \ba},
\label{frechet}
\eeq
as introduced by \citet{gilbert2018geometric}. It states that $\barbphi_\vp$ is the Fr\'echet mean of the maps $\barbphi$ for the $L_2$ distance. The equivalence follows from the equalities
\begin{align}
 \int \| \bm\psi (\ba,t) -  \barbphi (\ba,t) \|^2 \, \d \ba &=
\int \| \bm\psi (\ba,t) \|^2 - 2 \bm{\psi} (\ba,t) \bcdot \barbphi (\ba,t) + \| \barbphi (\ba,t) \|^2 \, \d \ba  \nonumber \\
&= \av{ \int \| \bm\psi (\ba,t) -  \bphi (\ba,t) \|^2 \, \d \ba} \nonumber \\
& \ \ +\int   \| \barbphi (\ba,t) \|^2 \, \d \ba -  \av{ \int \| \bphi (\ba,t) \|^2 \, \d \ba} 
\end{align}
using \eqref{meanmap}. The equivalence does not apply to time averaging because $\bm{\psi} (\ba,t) \bcdot \barbphi (\ba,t) \not= \av {\bm{\psi} (\ba,t) \bcdot \bphi (\ba,t)}$.

The minimization \eqref{min} is a standard problem in the theory of optimal transport \citep[e.g.][]{santambrogio2015optimal}. It leads to \citeauthor{brenier1991polar}'s \citep{brenier1991polar} polar factorization: the bare mean map can be written as the composition
\beq
\barbphi = \nabla \lambda \circ \barbphi_\vp
\label{polarfact}
\eeq
of the gradient of a convex scalar function $\lambda$ with the desired volume-preserving flow map. The characterization \eqref{polarfact} is central to the remainder of the paper. For completeness, we provide a formal derivation in appendix \ref{sec:polar}.

Given $\barbphi$, it is in principle possible to determine $\barbphi_\vp$ by solving either the minimization problem \eqref{min} or a Monge--Amp\`ere equation for the Legendre transform of $\lambda$ \citep{brenier1991polar}. For the purpose of computing Lagrangian means from simulation data, it is better to consider the time derivative of \eqref{polarfact} to obtain volume-preserving versions of \eqref{meanu2} and \eqref{zeromean}. 

Differentiating \eqref{polarfact} with respect to $t$ and using \eqref{meanu} gives
\beq
 \bnabla \lambda_t \circ \barbphi_\vp +  (\partial_t \barbphi_\vp \bcdot \bnabla) (\bnabla \lambda) \circ \barbphi_\vp  = \langle \bu \circ \bphi \rangle. \label{init_proj}
\eeq
The Lagrangian mean velocity $\barbu_\vp$ associated with $\barbphi_\vp$ satisfies
\beq
\partial_t \barbphi_\vp(\ba,t) = \barbu_\vp(\barbphi_\vp(\ba,t),t)
\label{meanu*}
\eeq
and is divergence free:
\beq
\bnabla \bcdot \barbu_\vp = 0.
\label{divu*}
\eeq
Using \eqref{meanu*} we rewrite \eqref{init_proj} as
\beq
\partial_t \bnabla \lambda  +  (\barbu_\vp \bcdot \bnabla) (\bnabla \lambda)  = \langle \bu \circ \bphi \rangle \circ \barbphi_\vp^{-1}. 
 \label{init_proj2}
\eeq
For the ensemble average we can freely move $\barbphi_\vp$ on the right-hand side within the average. Using the decomposition
\beq
\bphi = \bXi_\vp \circ \barbphi_\vp
\label{decomp*}
\eeq
then gives
\beq
\partial_t \bnabla \lambda  +  (\barbu_\vp \bcdot \bnabla) (\bnabla \lambda)  = \langle \bu \circ \bXi_\vp \rangle .
\label{lambdat}
\eeq
This equation is conveniently replaced by its divergence and curl, that is,
\begin{align} 
\partial_t \nabla^2 \lambda + \nabla \cdot ( (\barbu_\vp \bcdot \bnabla) (\bnabla \lambda) ) &= \nabla \cdot  \langle \bu \circ \bXi_\vp \rangle, \label{nabla2lam} \\
\nabla \times ( (\barbu_\vp \bcdot \bnabla) \nabla \lambda) &= \nabla \times \langle \bu \circ \bXi_\vp \rangle .
\label{curlbaru}
\end{align}
On the other hand, differentiating \eqref{decomp*} with respect to time leads to
\beq
\partial_t \bXi_\vp + (\barbu_\vp \bcdot \bnabla) \bXi_\vp = \bu \circ \bXi_\vp.
\label{Xi*}
\eeq

Eqs.\ \eqref{divu*}, \eqref{nabla2lam}, \eqref{curlbaru} and \eqref{Xi*} are key results of this section. They constitute a closed system that determines $\bu_\vp$, $\lambda$ and $\bXi_\vp$ from $\bu$. Specifically, \eqref{nabla2lam} and \eqref{Xi*} govern the joint evolution of $\lambda$ and $\bXi_\vp$ while  $\barbu_\vp$ is diagnosed from \eqref{divu*} and \eqref{curlbaru}. This system replaces the simpler GLM prescription that couples \eqref{Xi*} (without the dagger subscripts) and the relation $\barbu = \langle \bu \circ \bXi \rangle$ in  \eqref{meanu2}.

Once $\bXi_\vp$ is determined, the Lagrangian average of a quantity can be computed. For a scalar field $g(\bx,t)$, in particular, this average is defined by
\beq
\bar g_\vp(\barbphi_\vp(\ba,t),t) = \av{g(\bphi(\ba,t),t)}.
\label{avg1}
\eeq
(Again we keep the notation light by omitting a  superscript L for the Lagrangian average.)
For an ensemble average this reduces to
\beq
\bar g_\vp(\bx,t) = \av{g(\bXi_\vp(\bx,t),t)}.
\eeq

We conclude this section with three remarks. First, we  give an interpretation for the scalar field $\lambda$.
Averaging \eqref{Xi*} and comparing with \eqref{lambdat} shows that
\beq
\av{\bXi_\vp} = \nabla \lambda,
\label{avXi}
\eeq
assuming consistent initial conditions. Letting 
\beq
\lambda = \tfrac{1}{2}|\bx|^2 + \lambda',
\eeq
\eqref{avXi} is equivalent to
\beq
\av{\bxi_\vp} = \nabla \lambda'.
\label{lambdagrad}
\eeq
This condition, with $\lambda'$ determined by the condition that $\bu_\vp$ is divergence free, replaces GLM's condition $\av{\bxi}=0$. Thus, in the polar-factorization framework, the perturbation field has a non-zero mean given by the gradient of $\lambda'$.

The second remark concerns the difference between the bare GLM mean flow map $\barbphi$ and its volume-preserving approximation $\barbphi_\vp$. Provided that $\barbphi$ captures most of the dynamics in the sense that $\bXi$ is close to the identity (or, equivalently, $\bxi$ is small), the two maps are close together. We show this explicitly in appendix \ref{sec:smallamp} where we construct $\barbphi_\vp$ pertubatively assuming a velocity field of the form
\beq
\bu = \bu_0 + \eps \bu_1,
\label{pert}
\eeq
where $\bu_0$ is a fixed background velocity, $\bu_1$ a zero-mean perturbation velocity and $\eps \ll 1$. We establish  that $| \barbphi(\ba,t) - \barbphi_\vp(\ba,t)| = O(\eps^2)$ for arbitrarily large $t$. We also consider an alternative volume-preserving mean map proposed by \citet{sowa-robe10} and defined perturbatively (see also \citep{gilbert2018geometric,gilbertGeometricApproaches2024}).
This second volume-preserving mean map is also $O(\eps^2)$-close to $ \barbphi$. 

The final remark concerns a third volume-preserving mean map. Instead of approximating the GLM flow map $\barbphi$ by a volume-preserving flow map as we do for $\barbphi_\vp$, it is tempting to approximate the GLM mean velocity $\barbu$ by a divergence-free velocity field then use the corresponding flow map as volume-preserving mean map. The simplest construction takes the divergence-free velocity field as $\mathsf{P} \barbu$, the $L_2$-projection of  $\barbu$ on divergence-free vector fields. While this construction is seemingly appealing, we show in appendix \ref{sec:smallamp} that the resulting flow map does not remain close to $\barbphi$ for long times, so that it does not properly capture the mean dynamics.

\section{Numerical computation of Lagrangian means} \label{sec:exponential_mean}

\subsection{Exponential mean}

We now turn to the numerical computation of volume-preserving Lagrangian time averages. We focus on the exponential average
\beq
    \av{f}(t) =  \alpha \int_{-\infty}^{t} \e^{-\alpha(t -s)} f(s) \, \d s,
\eeq
where the parameter $\alpha>0$ is an inverse averaging time scale. The exponential average corresponds to the choice of kernel $k(t) = \alpha \,\e^{-\alpha t} H(t)$, with $H(t)$ the Heaviside function, in the general time average \eqref{timeav}. It has the advantage that $\av{f}(t)$ satisfies a differential equation, namely
\beq
\frac{\d \av{f}}{\d t} = \alpha (f - \av{f}).
\label{expav}
\eeq
This makes it possible to compute averages on-the-fly, as part of simulations that deliver $f(t)$ incrementally. \citet{minz2025efficient} exploit this property for the computation of Lagrangian averages and show it generalizes to $k(t)$ given by a sum of exponentials. \citet{baker2026offline} extends this to time-symmetric kernels by solving differential equations forward and backward in time for the offline computation of Lagrangian averages.

With the average \eqref{expav}, the right-hand side $\langle \bu \circ \bphi \rangle \circ \barbphi_\vp^{-1}$ of equation \eqref{init_proj2} governing the evolution of $\lambda$ can be evaluated explicitly. Integration by parts gives
\begin{align}
\langle \bu \circ \bphi \rangle(\ba,t) &= \alpha \int_{-\infty}^t \e^{-\alpha(t-s)} \partial_s \bphi(\ba,s) \, \d s \nonumber \\ 
&= \alpha \left( \bphi(\ba,t) - \alpha \int_{-\infty}^t \e^{-\alpha(t-s)}  \bphi(\ba,s) \, \d s \right) \nonumber \\
&= \alpha \left( \bphi(\ba,t) - \barbphi(\ba,t)\right). \label{ubar}
\end{align}
Using \eqref{polarfact} and \eqref{decomp*} then leads to
\beq
\langle \bu \circ \bphi \rangle \circ \barbphi_\vp^{-1} = \alpha (\bXi_\vp - \bnabla \lambda).
\eeq
Eq.\ \eqref{init_proj2} becomes
\beq
\partial_t \bnabla \lambda  +  (\barbu_\vp \bcdot \bnabla) (\bnabla \lambda)  = \alpha (\bXi_\vp - \bnabla \lambda).
\label{lamu}
\eeq
As in \S\ref{sec:vp}, it is convenient to take  the divergence and curl. This gives
\begin{subequations}
\begin{align}
\partial_t \nabla^2 \lambda + \bnabla \bcdot \left( (\barbu_\vp \bcdot \bnabla) \bnabla \lambda \right) &= \alpha ( \bnabla \bcdot \bXi_\vp - \nabla^2 \lambda)\quad \text{and} \label{nabla2lambda} \\
\bnabla \times \left((\barbu_\vp \bcdot \bnabla) \bnabla \lambda \right) &= \alpha \bnabla \times \bXi_\vp. \label{curlu*}
\end{align}
\label{2lambda}
\end{subequations}
These equations  should be solved together with the evolution equation \eqref{Xi*} for $\bXi_\vp$ and the divergence-free condition \eqref{divu*} for $\barbu_\vp$. Eq.\ \eqref{nabla2lambda} governs the evolution of $\nabla^2 \lambda$ and, on inverting the Laplacian, $\lambda$. Eqs.\ \eqref{divu*} and \eqref{Xi*} determine $\barbu_\vp$. In an on-the-fly implementation, this system of equations is solved at the same time as dynamical equations for the instantaneous velocity $\bu$. With $\bXi_\vp$ and $\barbu_\vp$ determined, the volume-preserving Lagrangian average of any field can be computed by solving an additional PDE. For a scalar field $g(\bx,t)$, this takes the form
\beq
\partial_t \bar g_\vp + \barbu_\vp \bcdot \bnabla \bar g_\vp =  \alpha \left( g \circ \bXi_\vp - \bar g_\vp \right)
\label{barg*}
\eeq
as follows from differentiating the definition
\beq
\bar g_\vp(\barbphi_\vp(\ba,t),t) = \alpha \int_{-\infty}^t \e^{-\alpha(t-s)} g(\bphi(\ba,s),s) \, \d s
\eeq
with respect to $t$.

Volume-preserving Lagrangian averaging based on polar factorisation is not limited to the exponential mean but can be adapted to various types of filters, e.g.\ those proposed in Refs.\ \cite{baker2024lagrangian,minz2025efficient}. We illustrate this point by developing a volume-preserving version of the Butterworth filter in appendix \ref{app:butter}.


\subsection{Application to a two-dimensional incompressible flow} \label{sec:2d}

We implement the system above for solutions of the two-dimensional incompressible Euler equations.  
This system does not admit fast wave motion. We use it solely to compare the impact of standard and volume-preserving Lagrangian averaging on the slow vortical flow. In vorticity--streamfunction formulation, the Euler equations read
\begin{equation}
    \partial_t \zeta + \bu \bcdot \bnabla \zeta = 0 \quad \textrm{with} \quad \bu = \bnabla^\perp \psi
   \quad \textrm{and} \quad \nabla^2 \psi = \zeta,
    \label{eq: 2dNS}
\end{equation}
where $\zeta$ is the vorticity, $\psi$ the streamfunction and $\bnabla^\perp=(-\partial_y, \partial_x)$.

We solve the Euler equations \eqref{eq: 2dNS}  in a $2\pi \times 2 \pi$ periodic domain using a modified version of the pseudospectral code in Ref.\ \citep{kafiabad2023computing}. We use $256^2$  grid points and Fourier modes, and an RK4 integrator with time step $\Delta t =  0.005$.  For numerical stability, we employ 2/3 dealiasing and we add hyperviscous dissipation to the vorticity equation by multiplying the Fourier transform $\hat{\zeta}(\bk,t)$ of the vorticity by $\exp(-\kappa |\bk|^8 \Delta t)$ at each time step. We take $\kappa = 2.6 \times 10^{-14}$. 

Together with the Euler equations, we solve the equations required for the computation of Lagrangian averages. For  volume preserving averaging, these consist of \eqref{Xi*}, \eqref{2lambda} and \eqref{barg*}. To deal with periodic fields, we use $\bxi_\vp$ instead of $\bXi_\vp$ and $\lambda' = \lambda - \tfrac{1}{2} \|\bx\|^2$ instead of $\lambda$. We enforce the divergence-free constraint for $\barbu_\vp$ by means of a streamfunction, letting
\beq
\barbu_\vp = \bnabla^\perp \bar \psi_\vp.
\eeq
With these changes, \eqref{Xi*} becomes
\beq
\partial_t \bxi_\vp + \barbu_\vp \bcdot \bnabla \bxi_\vp = \bu \circ \bXi_\vp - \barbu_\vp,
\label{cc}
\eeq
and \eqref{2lambda} become
\begin{subequations} \label{lampsi}
\begin{align}
\partial_t \nabla^2 \lambda' + \bnabla \bcdot \left( (\barbu_\vp \bcdot \bnabla) \bnabla \lambda' \right) &= \alpha (  \bnabla \bcdot \bxi_\vp - \nabla^2 \lambda'), \label{aa} \\
\nabla^2 \bar \psi_\vp + \bnabla \times \left((\barbu_\vp \bcdot \bnabla) \bnabla \lambda' \right) &= \alpha \bnabla \times \bxi_\vp. \label{bb}
\end{align}
\end{subequations}
At each time step, we update $\bxi_\vp$ and $\lambda'$ using \eqref{cc} and \eqref{aa} then compute  $\bar \psi_\vp$ by solving \eqref{bb} using the iteration
\beq
\nabla^2 \bar \psi^k_\vp  = \alpha \bnabla \times \bxi_\vp - \bnabla \times \left((\barbu_\vp^k \bcdot \bnabla) \bnabla \lambda'\right),
\eeq
where $k=0,\, 1, \cdots$ denotes the iterate and $\barbu_\vp^k = \bnabla^\perp \bar \psi_\vp^k$. We focus on the Lagrangian average $\bar \zeta_\vp$ of the vorticity, computed from \eqref{barg*} with $g =\zeta$. We use bilinear interpolation to evaluate $\bu$ and $g$ at position $\bXi_\vp(\bx,t) = \bx + \bxi_\vp(\bx, t)$ in \eqref{barg*} and  \eqref{cc}.
For numerical stability,  we find it necessary to add to \eqref{cc} the same hyperviscous dissipation as in the vorticity equation. 
For comparison, we also compute the standard GLM average $\bar \zeta$ of the vorticity. For this, we solve the bare versions of \eqref{barg*} and  \eqref{cc} with $\barbu = \alpha \bxi$, as follows from \eqref{ubar} \citep[see also][]{minz2025efficient}.

    \begin{figure}
        \includegraphics[width=1.005\linewidth]{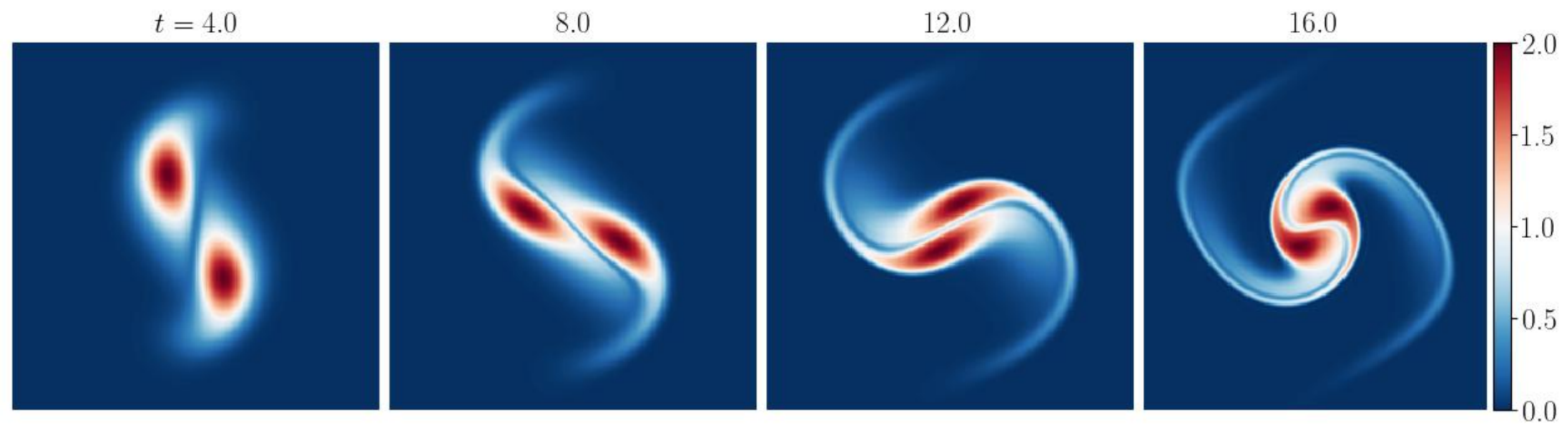}\\
        \includegraphics[width=1.005\linewidth]{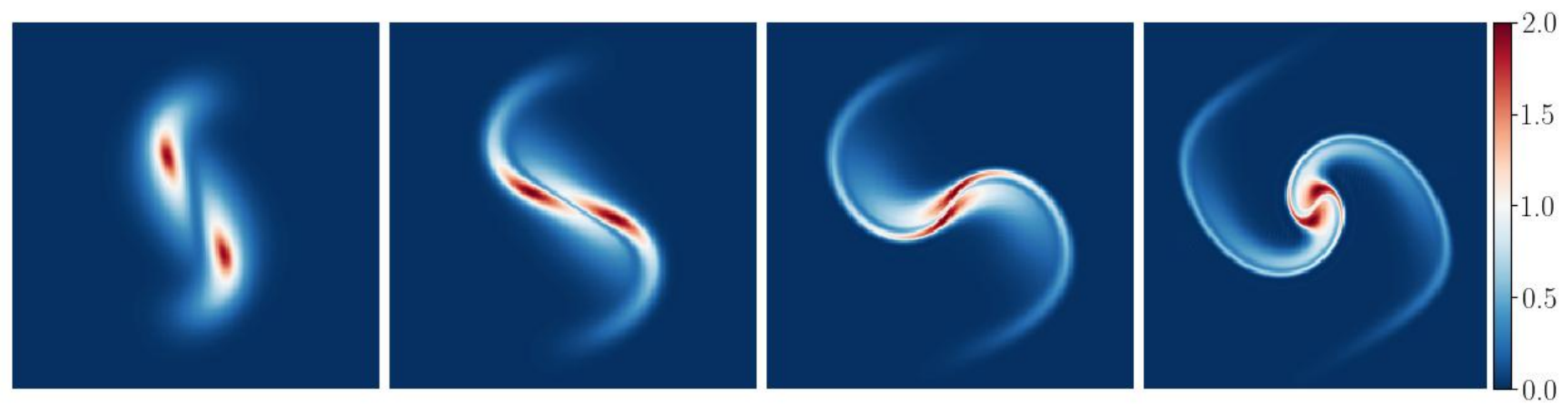} \\
        \includegraphics[width=1.005\linewidth]{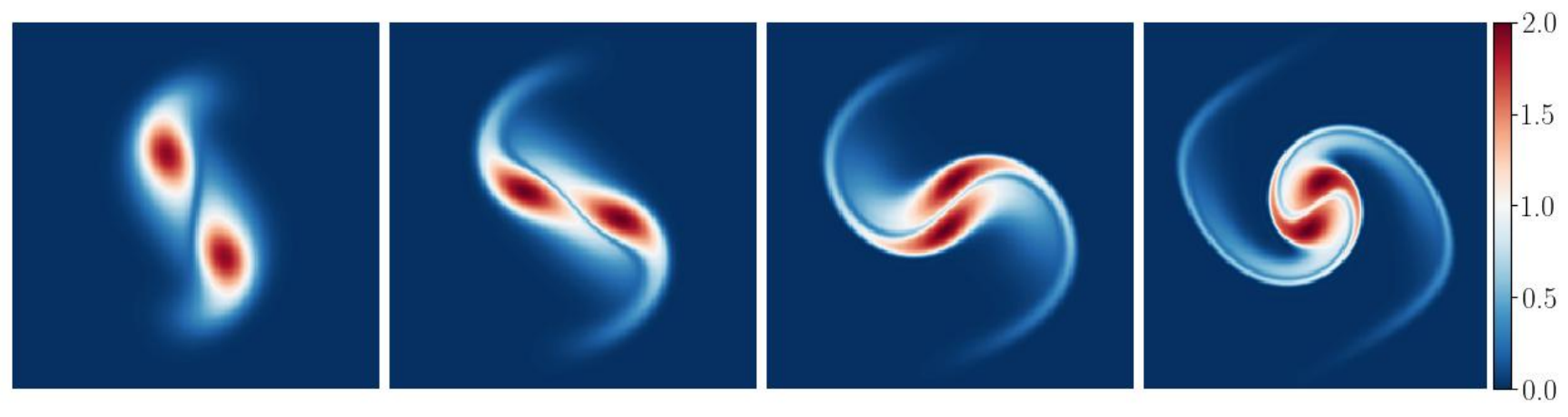} 

        \includegraphics[width=1.005\linewidth]{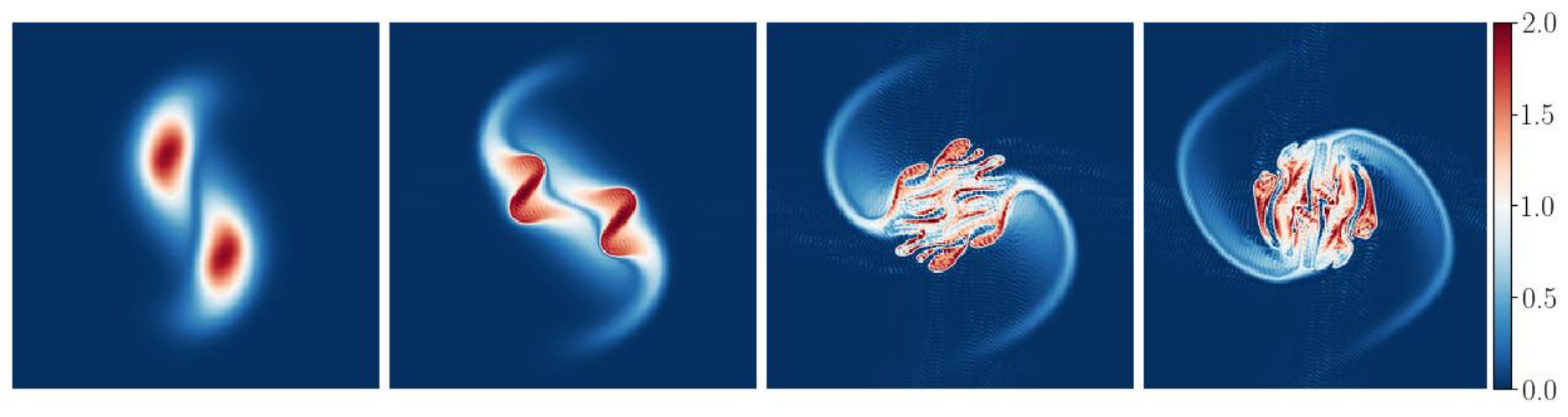} 
        
        \caption{Vorticity  in a simulation of two-dimensional incompressible fluid at $t =4, 8,12$ and $16$: instantaneous vorticity $\zeta$ ($1^{\text{st}}$ row), GLM mean $\bar {\zeta}$ ($2^{\text{nd}}$ row), volume-preserving mean $\bar \zeta_\vp$  ($3^{\text{rd}}$ row) and  mean obtained using the divergence-free projection of the GLM mean velocity ($4^{\text{th}}$ row). We use an exponential average with $\alpha = 0.5$. An animation of the vorticity fields is available at \url{https://abhijeet-minz.github.io/animations.html}.} \label{merger}
    \end{figure}

The top row of figure \ref{merger} shows the evolution of the vorticity $\zeta$ for an initial condition that consists of two well-separated vortices, specifically
\beq
\zeta(\bx,t=0) = 2 \left(
\e^{-2.5 \left(x^2 + \left(y + \pi/3\right)^2 \right)}
+ \e^{-2.5 \left(x^2 + \left(y - \pi/3 \right)^2 \right)} 
\right).
\eeq
The dynamics are familiar: the vortices orbit the central point $(0,0)$, deform and undergo vortex merger. The second and third  rows show the  Lagrangian averages $\bar \zeta$ and $\bar \zeta_\vp$ of the vorticity obtained with the exponential filter with inverse time scale $\alpha = 0.5$. To mitigate the effect of the phase delay that affects the exponential filter, we offset time by the zero-frequency phase delay $\av{t} = \alpha^{-1} = 2$. 

The bare Lagrangian averaging exhibits the expected divergence effect: the vortices in the Lagrangian mean field $\bar \zeta$ have much decreased areas compared with those in the instantaneous field $\zeta$. This is successfully corrected by the volume-preserving average. The averaging distorts the vorticity field, primarily because the frequency-dependent phase shift it introduces is only partially compensated by the time offset. Thus, while the slowly evolving filaments at the periphery of the vortex merger are in phase in $\zeta$ and $\bar \zeta_\vp$, the faster evolving cores are not. 

\begin{figure}
\begin{center}
     \includegraphics[width = 0.5\textwidth]{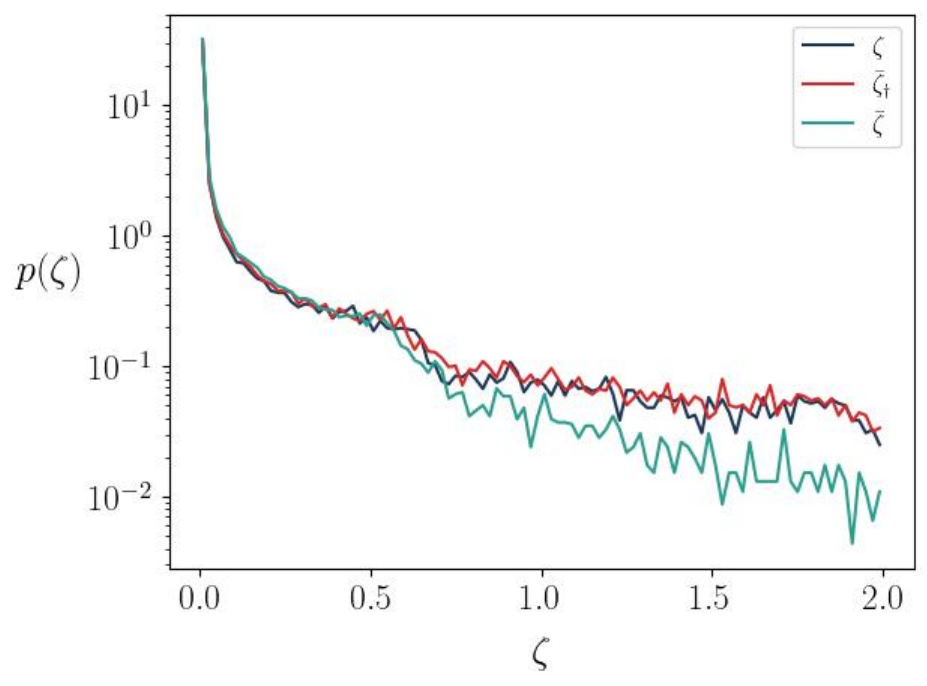} \\
  \caption{Comparison of the pdfs of $\zeta$, $\bar \zeta$ and $\bar \zeta_\vp$ in the simulation of figure \ref{merger} at $t=20$.}
 \label{fig:pdf_2dNS}
 \end{center}  
\end{figure}

The bottom row of figure \ref{merger} gives a finite-amplitude illustration of the difficulties that arise with the simple replacement of the bare mean velocity $\barbu$ by its divergence-free projection $\mathsf{P} \barbu$ discussed for small-amplitude perturbations in appendix \ref{sec:smallamp}. With this choice, the mean flow map does not stay close to the actual flow map; equivalently, the perturbation map $\bXi$ does not stay close to the identity map. As a result, the mean vorticity field develops small-scale features unrelated to the instantaneous dynamics.

A complementary view of the divergence effect and its correction by  volume-preserving averaging is offered in figure \ref{fig:pdf_2dNS}. The figure shows the probability density function (pdf) of $\zeta$, $\bar \zeta$ and $\bar \zeta_\vp$ at $t = 20$. The pdf of $\zeta$ is time independent up to numerical effects (truncation error and dissipation). The pdfs of $\bar \zeta$ and $\bar \zeta_\vp$ are also time independent after an initial adjustment phase. The depletion of high values of $\bar \zeta$ reflects the decrease in area of the vortices noted in figure \ref{merger}. The pdf of $\bar \zeta_\vp$, in contrast, is almost identical to that of $\zeta$, confirming the effectiveness of the volume-preservation algorithm.




\subsection{Application to a rotating shallow-water flow}  \label{sec:SW}

In this section, we apply Lagrangian averaging to a two-time-scale flow, namely a rapidly rotating shallow-water flow that combines slow turbulent vortical motion with a fast mode-1 Poincar\'e wave. 
Such a flow is used in Refs.\ \cite{kafiabad2022grid,baker2024lagrangian,minz2025efficient} to demonstrate the capabilities of different PDE-based averaging methods. 

At first sight, the divergence effect that we address in the present paper does concern the shallow-water model. In this model, the (two-dimensional, depth-averaged) velocity field is divergent because of the free surface, so there appears to be no reason to insist that the mean flow should preserve areas. However, in the rapidly rotating, small-Rossby-number regime that we consider, the vortical part of the flow is approximately in geostrophic balance. This implies that the corresponding velocity is approximately divergence free. More precisely, the Lagrangian mean velocity is approximately geostrophically balanced and divergence free \cite[see][and references therein]{kafiabad2021wave}. For sufficiently small Rossby number, we expect the divergence effect associated with the definition of the Lagrangian mean flow to dominate the small physical divergence associated with finite Rossby number. It is therefore sensible to impose exact area preservation in order to eliminate the divergence effect. 

We carry out a simulation using the rotating shallow water model, as given in \cite{minz2025efficient}. The non-dimensional momentum and continuity equations are \citep{zeitlin2018geophysical}
\begin{subequations}  \label{eq:shallow water}
\begin{align}
           \partial_t{\bu} + \bu \bcdot \bnabla \bu + Ro^{-1} \hat{\bm{z}} \times \bu &= - Fr^{-2} \bnabla h, \label{eq:shallow water (a)} \\
        \partial_t{h} + \bnabla \bcdot (h \bu) &= 0, 
\end{align}
\end{subequations}
where $Ro = U/(fL)$ is the Rossby number and $Fr = U/\sqrt{gH}$ is the Froude number. Here $U$, $L$ and $H$ are the reference velocity, length and depth used for the non-dimensionalization. 

We initialise the model with the same initial condition as  in \cite{kafiabad2023computing} and \cite{minz2025efficient}, superimposing a geostrophic turbulent flow (obtained by prior solution of an incompressible two-dimensional Navier–Stokes model for the velocity and geostrophic balance for the height field) and a right-travelling mode-$1$ Poincar\'e wave. 
We take the root-mean square velocity of the geostrophic flow as reference velocity $U$ and the length scale of the first Fourier mode as reference length $L$ so that the doubly periodic domain is $[0 ,2\pi]^2$. The Rossby and Froude number of the geostrophic flow are $Ro = 0.1$ and $Fr = 0.5$. With these parameters, the frequency $\omega = (Ro^{-2} + Fr^{-2})^{{1}/{2}}$ of the mode-$1$ Poincaré wave is $\omega = 10.2$ corresponding to a period of $0.62$ time units.
The wave fields in the absence of flow are given by
\begin{equation}
    u' = a\ \cos(x - \omega t), \quad v' = \frac{a}{\omega Ro} \ \sin(x - \omega t), \quad h' = \frac{a}{\omega} \ \cos(x - \omega t). \label{eq:wave_amp}
\end{equation}
At the initial time $t=0$, we add these to the geostrophic flow field. We take the amplitude $a=-1$ so that the maximum wave velocity is as large as the geostrophic flow root-mean-square velocity.

\begin{figure}
        \includegraphics[width=1.00\linewidth]{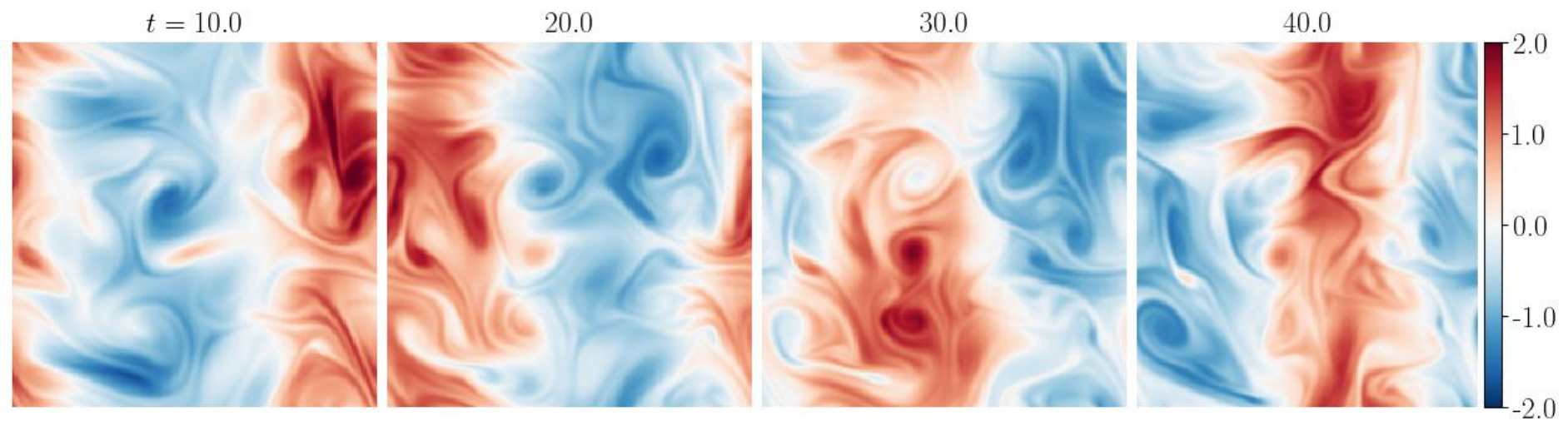}\\
        \includegraphics[width=1.00\linewidth]{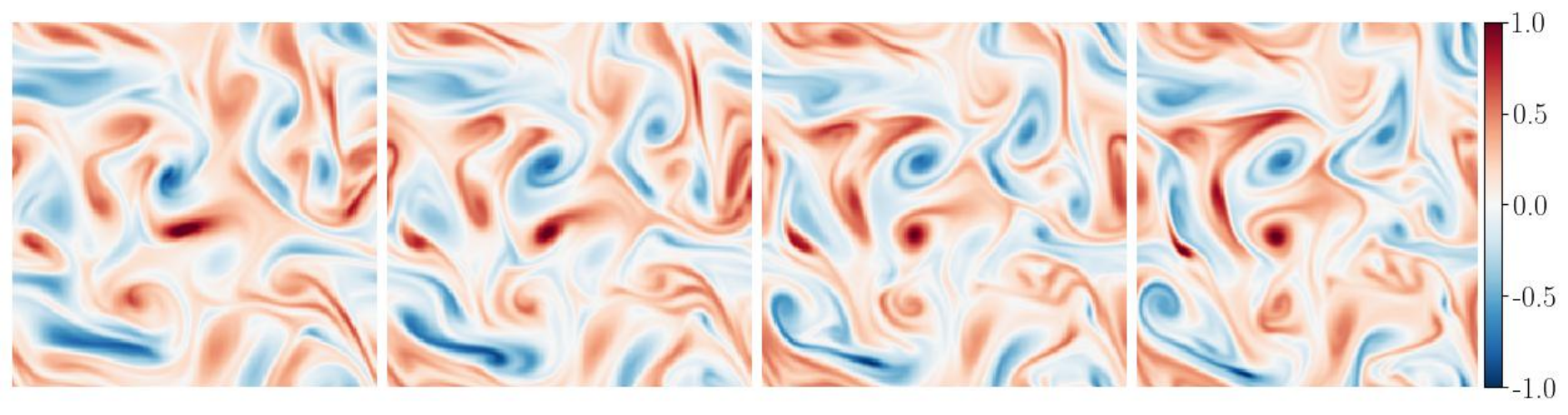} \\
        \includegraphics[width=1.00\linewidth]{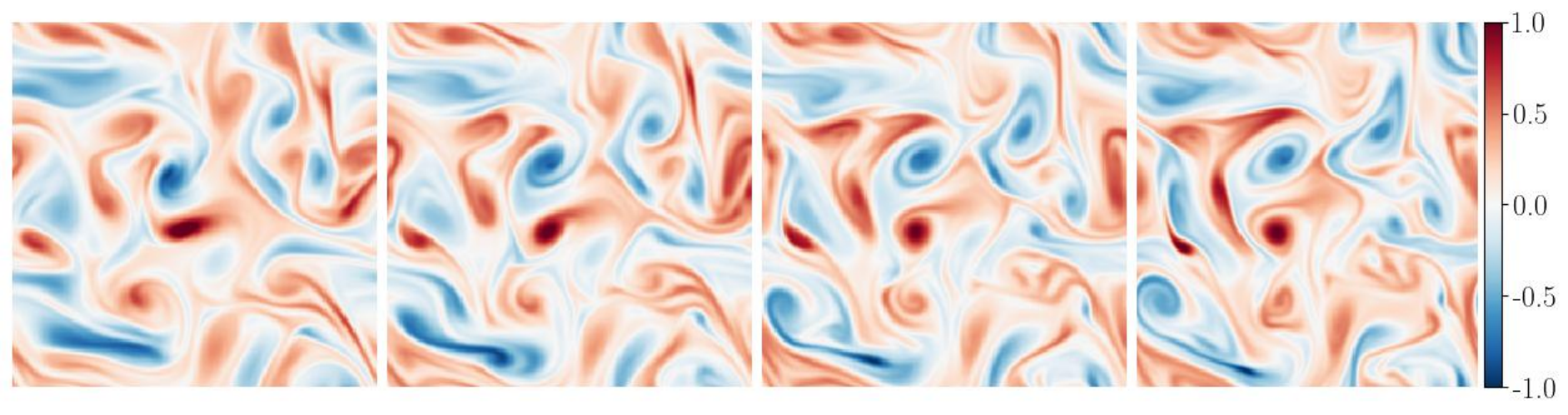}      
        \caption{Vorticity field at four successive times in the shallow-water simulation of \S\ref{sec:SW}: the instantaneous vorticity $\zeta$ (top row) is compared with its bare and volume-preserving exponential Lagrangian averages $\bar\zeta$ and  $\bar{\zeta}_\vp$ (second and third rows). An animation of the vorticity fields is available at \url{https://abhijeet-minz.github.io/animations.html}.}
        \label{fig:rsw}
    \end{figure}

To obtain the volume preserving Lagrangian mean of a scalar field, we solve \eqref{barg*} along with the dynamical equations \eqref{eq:shallow water}, equation \eqref{cc}  for $\bxi_\vp$ and equations \eqref{lampsi} for $\lambda'$ and $\psi_\vp$. We use a pseudospectral implementation, identical to that of \S\ref{sec:2d} for \eqref{cc} and \eqref{lampsi}.
For numerical stability, we apply the same dissipation scheme to \eqref{eq:shallow water (a)} and \eqref{cc} as in \S\ref{sec:2d}.

Figure \ref{fig:rsw} shows the evolution of the vorticity $\zeta$ and of its bare and volume-preserving averages $\bar \zeta$ and $\bar \zeta_\vp$. The instantaneous vorticity is dominated by the large-scale wave signal, but vortical structures are visible as smaller scale modulations. Much of the wave signal is eliminated by the averaging, leaving similar-looking fields $\bar \zeta$ and $\bar \zeta_\vp$. The vortices in $\bar \zeta$ are smaller than those in $\bar \zeta_\vp$ and, though this is less obvious to discern, in $\zeta$. Vorticity (as opposed to potential vorticity) is not materially conserved in shallow water, so the area of vortices change in time. Nonetheless we argue that the area changes in $\bar \zeta$ are the result of the divergence effect rather than of a physical divergence. 

\begin{figure}
\begin{center}
     \includegraphics[width = 1\textwidth]{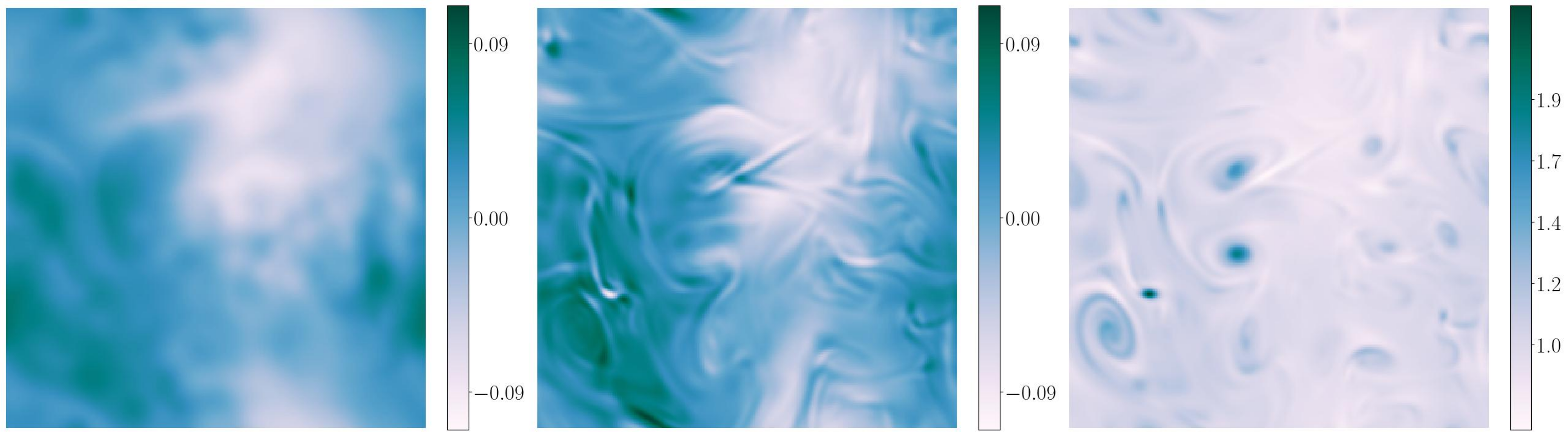} \\
  \caption{Divergence of the Eulerian and bare Lagrangian mean velocities, $\nabla \bcdot \langle \barbu \rangle$ (left panel) and $ \nabla\bcdot \barbu$ (middle panel) at the end of the shallow-water simulation ($t=40$). The averages use the exponential filter with $\alpha=0.5$.  The right panel shows the Jacobian $ |\partial \bXi / \partial \bx|$ at the same time.}
 \label{fig:div}
 \end{center}  
\end{figure}

To support this, we show in figure \ref{fig:div} the divergences $\nabla \cdot \av{\bu}$ and $\nabla \cdot \bar{\bu}$ of the Eulerian and bare Lagrangian mean velocities. The bare Lagrangian mean $\barbu$ for the exponential mean  is computed from the relation $\barbu = \alpha (\bXi-\bx)$ (which is shown in \cite{minz2025efficient} or can be deduced from \eqref{lamu} by setting $\lambda = |\bx|^2/2$). 
Both $\nabla \cdot \av{\bu}$ and $\nabla \cdot \bar{\bu}$ are dominated by a wave signal that results from the limitations of exponential averaging as a low-pass filter. The slow signal, correlated with the vortical structure in figure \ref{fig:rsw}, is much larger in 
$\nabla \cdot \bar{\bu}$, suggesting it results from the averaging  rather than from a physical process. We also show in figure \ref{fig:div} (right panel) the Jacobian field $|{\partial \bXi(\bx,t)}/{\partial \bx}|$. Since
\beq
 \left| \frac{\partial \bXi(\bx,t)}{\partial \bx} \right| =  \left| \frac{\partial \bphi(\ba,t)}{\partial \ba} \right|   \left| \frac{\partial \barbphi(\ba,t)}{\partial \ba} \right|^{-1} 
\eeq
this Jacobian is the ratio of the height changes associated with the compressibility of $\barbphi$ and $\bphi$, respectively. The compressibility of $\bphi$ is weak because $\nabla \cdot \bu$ is dominated by an oscillatory wave component that has little time-integrated effect.
The large values in $|{\partial \bXi(\bx,t)}/{\partial \bx}|$, up to $2$, are dominated by the compressibility of $\barbphi$ induced by the divergence effect.

\begin{figure}
\begin{center}
     \includegraphics[width = 1.005\textwidth]{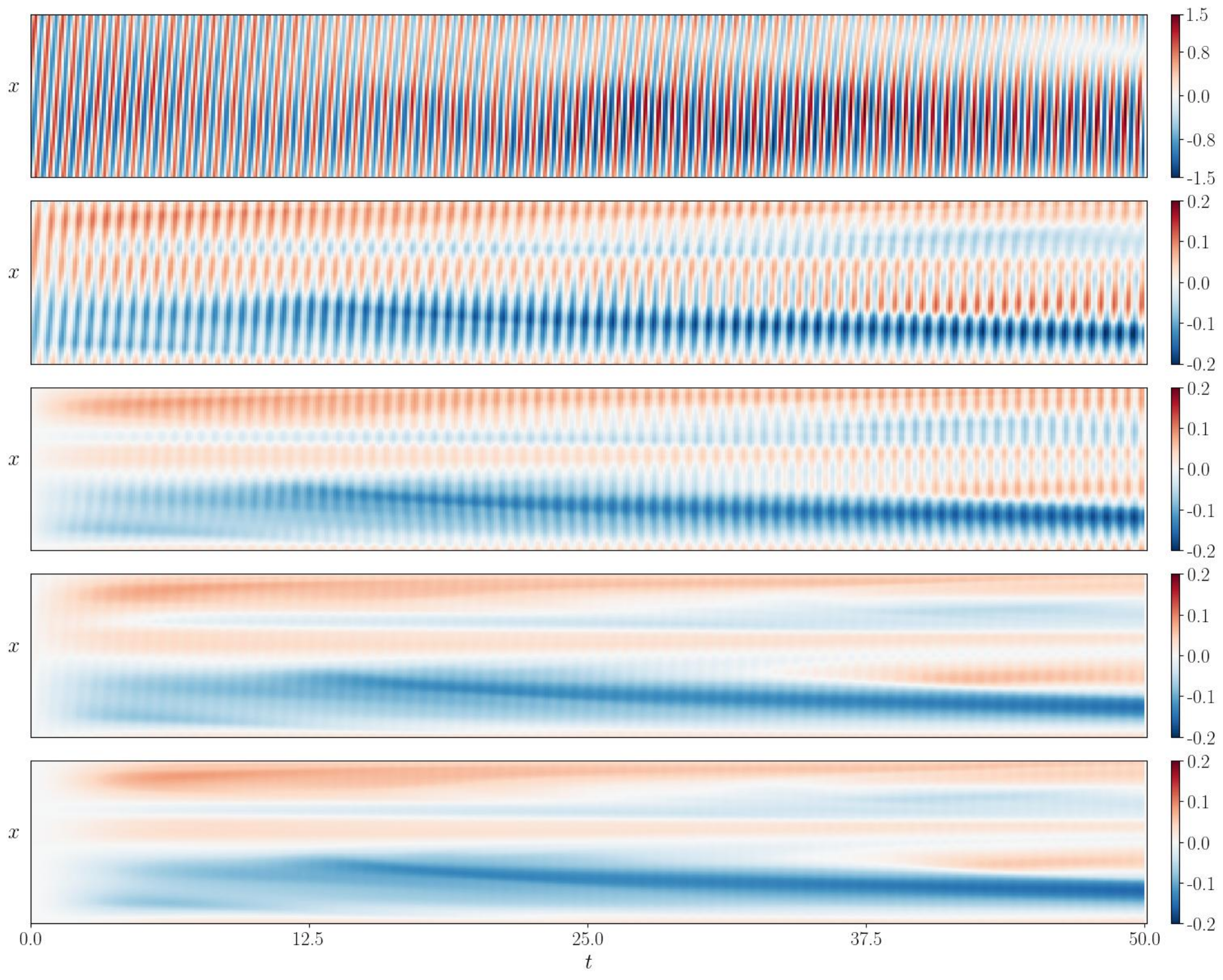} \\
  \caption{First component of the velocity in the shallow-water simulation as a function of $t$ and $x$ for $y=0.24$: the  instantaneous velocity $u$ (first row) is compared with the Lagrangian mean velocities obtained using the bare exponential filter (second row), the volume-preserving exponential filter (third row), the bare 2nd-order Butterworth filter (fourth row), and the volume-preserving 2nd-order Butterworth filter (fifth row).}
 \label{fig:u}
 \end{center}  
\end{figure}

Figure \ref{fig:u} shows the first components $u$, $\baru$ and $\baru_\vp$ of the velocity and of its Lagrangian means at fixed $y=0.24$ as functions of $t$ and $x$. The instantaneous $u$ (top row) is dominated by the fast Poincar\'e wave. The bare Lagrangian mean $\baru$ primarily captures the slow dynamics of the simulation, but a substantial wave signal remains, reflecting the limitation of the exponential mean as a low-pass filter. Remarkably, the wave signal is much reduced in the volume-preserving version $\baru_\vp$ (third row) obtained from \eqref{2lambda}.  We attribute this improvement to the fact that the Poincar\'e wave is divergent and  partially filtered from $\barbu_\vp$ by the divergence-free constraint $\nabla \cdot \bu_\vp = 0$. We find the same improvement between the bare and volume-preserving Lagrangian averages when the exponential filter is replaced by the 2nd-order Butterworth filter (fourth and fifth rows of figure \ref{fig:u}). The benefit of volume preservation combines with the better low-pass filtering of the Butterworth filter to produce a Lagrangian mean velocity $\baru_\vp$ that is almost perfectly free of fast oscillations.

\section{Discussion}

This paper develops and implements a method for the computation of Lagrangian averages referred to mean positions that are related to the initial positions of particles by a volume-preserving mean flow map $\barbphi_\vp$. Equivalently, the Lagrangian mean velocity $\barbu_\vp = \partial_t \barbphi_\vp \circ \barbphi_\vp^{-1}$ is divergence free: $\nabla \cdot \barbu_\vp = 0$. For incompressible fluids, this also means that the perturbation map $\bXi_\vp$ between mean and actual positions preserves volume. 

The volume-preservation properties are in contrast with the standard GLM for which the mean and perturbation  maps $\barbphi$ and $\bXi$ do not preserve volume and the Lagrangian mean velocity is non-divergent, $\nabla \cdot \barbu \not= 0$. Our construction is nonetheless anchored to standard GLM since $\barbphi_\vp$ is chosen as the volume-preserving map closest to $\barbphi$ or, equivalently, as the volume-preserving factor in the polar factorization of $\barbphi$. Note however that $\barbphi_\vp$ can be alternatively defined as the Fr\'echet mean of an ensemble of flow maps, without explicit reference to the GLM mean map (see \eqref{frechet}). 

The polar-factorization characterization we use relies on the (arbitrary) choice of the $L_2$-norm to measure the closeness of  $\barbphi_\vp$ to $\barbphi$. In principle, other $L_p$-norms could be chosen, but the corresponding optimal-transport theory is considerably more complicated.  Alternatively, as proposed in Ref.\ \citep{gilbert2018geometric}, it is possible to replace the pointwise $L_2$-norm by a more intrinsic measure of distance between maps, namely the distance that arises naturally in the interpretation of flow maps  as points in a Lie group  with right-invariant metric \citep[e.g.][]{arnold1998topological}.  
Turning the resulting abstract definition of the mean map (termed ``geodesic mean'' in \citep{gilbert2018geometric}) into a practical algorithm is a formidable challenge, however. The volume-preserving mean of \citet{sowa-robe10} (termed ``glm'', see \citep{gilbert2018geometric}), which is defined perturbatively assuming small-amplitude perturbations, is similarly challenging.
With this in mind, we contend that the polar-factorization definition of a volume-preserving mean map offers the most convenient way of overcoming the divergence effect of GLM in numerical Lagrangian averaging. 

Our simulation results suggest that the divergence effects can strongly distort the slow dynamics. This may not have been fully appreciated in the past because most studies of GLM assume small-amplitude perturbations and track the slow dynamics over relatively short times. Over multiple eddy times scales, however, the divergence effect leads to a significant contraction of vortices and, conversely, significant expansion of hyperbolic regions. This occurs even though the divergence of the Lagrangian mean velocity is weak because the process of volume or area change is multiplicative. The distortion of the slow dynamics is problematic when Lagrangian averaging is used to separate fast waves from slow flows by substracting full and averaged fields \citep[e.g.][]{baker2024lagrangian}. It would be worthwhile investigating how the adoption of the volume-preserving mean flow of this paper improves this separation. 

This paper focuses entirely on the kinematics of GLM and its volume-preserving variant. Key advantages of GLM over Eulerian averaging lie however in the structure of its averaged dynamical equations. An important question is therefore whether this structure persists for volume-preserving Lagrangian averaging. The answer is yes. As \citet{gilbert2018geometric,gilbertGeometricApproaches2024} emphasise, much of the structure of the GLM equations is independent of the specific choice made for the Lagrangian mean map. The Lagrangian mean momentum equation, in particular, is only affected through changes in the specific form of the pseudomomentum, with extra terms arising because $\langle \bxi \rangle \not= 0$. Whether the associated complications outweight the benefits brought about by volume preservation may be a matter ot taste.

\vspace{1\baselineskip} 
\label{SupMat}Supplementary movies are available at \url{https://abhijeet-minz.github.io/animations.html}

\vspace{1\baselineskip} 

\label{SupMat2}The code developed for the numerical applications in \S\ref{sec:2d} and \S\ref{sec:SW} and  is  openly available \citep{MinzZenodo}.

\vspace{1\baselineskip} 

We are grateful to H. A. Kafiabad and C. Maitland-Davies for useful discussions.

\vspace{1\baselineskip} 
The authors report no conflict of interest.

\appendix

\section{Rotating particle} \label{app:rigid}

Consider a two-dimensional fluid in rigid rotation with angular velocity $\Omega(t)$.
The position of a particle, $z(t) = x(t) + \i y(t)$ in complex notation, satisfies
\beq
\dot z = \i \Omega z
\eeq
and is given by 
\beq
z(t) = \e^{\i \theta(t)} c,
\eeq
where
\beq
 \theta(t) = \int_0^t \Omega(s) \, \d s.
\eeq
The initial position $c = a + \i b$ can be interpreted as a label for the particle. The corresponding Lagrangian mean position is
\beq
\bar z(t) = \chi(t) c, \quad \textrm{where} \ \ \chi(t) = \av{\e^{\i \theta(t)}}.
\label{barz}
\eeq

We focus on time averages of the form \eqref{timeav}.
With the additional assumption $k(t) \ge 0$, the triangle inequality implies that $|\chi| \le 1$, hence 
\beq
| \bar z(t)| = | \chi| |c| \le |c| = |z(t)|. 
\eeq
This shows that averaging contracts trajectories.
The Lagrangian mean velocity 
\beq
\bar u + \i \bar v = \dt{}{t} \ln \chi  \, \bar z
\eeq
is divergent,
\beq
\bar u_x + \bar v_y = 2 \, \Re \dt{}{t} \ln \chi  = 2 \dt{}{t} \ln |\chi|  \not=0
\eeq
unless $\dot \theta = \mathrm{const}$. 

In this paper we propose a definition of the mean flow map that replaces the straightforward coordinate-wise mean of GLM by the closest volume-preserving flow map. For rigid-body rotation, this amounts to replacing $\bar z(t)$ in \eqref{barz} by the nearest point on the circle of radius $|z(t)| = |c|$, that is, by
\beq
\bar z_\vp(t) = \frac{\chi}{|\chi|} c = \e^{\i \arg \chi} \, c. 
\eeq
This represents a rigid-body rotation. 
The velocity
\beq
\bar u_\vp + \i \bar v_\vp = \i \dt{}{t} \arg \chi \, \bar z_\vp
\eeq
is divergence-free.

\section{Polar factorization} \label{sec:polar}

We give a formal argument for the polar factorization \eqref{polarfact}. First observe that
\beq
\int \| \bm{\psi}(\ba,t) - \barbphi(\ba,t)\|^2 \, \d \ba = - 2 \int \bm{\psi}(\ba,t) \cdot \barbphi(\ba,t) \, \d \ba + \mathrm{const}, 
\label{cost}
\eeq
where the constant is independent of $\bm{\psi}$ because
\beq
\int \| \bm{\psi}(\ba,t) \|^2 \, \d \ba = \int \|\bx\|^2 \, \d \bx,
\eeq
using that $\bm{\psi}$ preserves volume. We next consider the maps
$\bm{\psi}$ as near-identity, volume-preserving perturbations of the extremum $\barbphi_\vp$. Near-identity perturbations can be expressed in terms of a generating vector field $\bm{w}$ as
\beq
\e^{\bm{w} \bcdot \bnabla} \bx = \bx  + \bm{w}(\bx,t) +  \tfrac{1}{2} (\bm{w} \bcdot \bnabla \bm{w}) (\bx,t) + \cdots.
\eeq
The divergence-free condition $\bnabla \bcdot \bm{w}=0$ ensures that these perturbations preserve volume. With $\bm{\psi} = \e^{\bm{w} \bcdot \bnabla} \circ \barbphi_\vp$, this gives
\beq
\bm{\psi}(\ba,t) =  \barbphi_\vp(\ba,t) +  \bm{w}(\barbphi_\vp(\ba,t),t) + \tfrac{1}{2} (\bm{w} \bcdot \bnabla \bm{w}) (\barbphi_\vp(\ba,t),t) + \cdots.
\eeq

With this expression and \eqref{cost}, the condition that $\bm{\psi} = \barbphi_\vp$ minimises the left-hand side of  \eqref{cost} reads
\begin{align}
0 &\ge  \int \bm{\psi}(\ba,t) \cdot \barbphi(\ba,t) \, \d \ba - \int \barbphi_\vp(\ba,t) \cdot \barbphi(\ba,t) \, \d \ba \nonumber \\
&= \int  \bm{w}(\barbphi_\vp(\ba,t),t)  \bcdot \barbphi(\ba,t)  \, \d \ba  + \tfrac{1}{2} \int  (\bm{w} \bcdot \bnabla \bm{w}) (\barbphi_\vp(\ba,t),t) \bcdot \barbphi(\ba,t) \, \d \ba \nonumber  \\
&= \int  \bm{w}(\bx,t)  \bcdot (\barbphi \circ \barbphi_\vp^{-1} (\bx,t)) \, \d \bx  + \tfrac{1}{2} \int  (\bm{w} \bcdot \bnabla \bm{w}) (\bx,t) \bcdot ( \barbphi \circ \barbphi_\vp^{-1} (\bx,t))  \, \d \bx,
\label{ineq}
\end{align}
on changing integration variable from $\ba$ to $\bx = \barbphi_\vp(\ba,t)$ and omitting $O(\| \bm{w} \|^3)$ terms. 
For the inequality to hold,
the first integral on the last line of \eqref{ineq} must vanish 
for all divergence-free $\bm{w}$. This implies that 
\beq
\barbphi \circ \barbphi_\vp^{-1} = \nabla \lambda
\label{polar2}
\eeq
for some $\lambda$, as follows from the orthogonality of the Helmholtz decomposition  \citep[e.g.][]{girault1986finite}. Rearranging gives the polar factorization \eqref{polarfact}. Introducing \eqref{polar2} into the second integral and integrating by parts gives
\beq
\int \bm{w}(\bx,t) \bcdot \bnabla \bnabla \lambda(\bx,t) \bcdot \bm{w}(\bx,t) \, \d \bx \ge 0,
\eeq
i.e.\ convexity of $\lambda$.

\section{Small-amplitude perturbations} \label{sec:smallamp}

We now obtain an approximation to the volume-preserving Lagrangian mean velocity $\barbu_\vp$ for 
a velocity field of the form \eqref{pert}.
 The velocity is divergence free such that $\bnabla \bcdot \bu_0 = \bnabla \bcdot \bu_1 = 0$. We use an ensemble average, with $\av{\bu_1} = 0$.

We solve \eqref{divu*}, \eqref{lambdat} and \eqref{Xi*} perturbatively. We expand
\begin{subequations}
\begin{align}
\bXi_\vp &= \bx + \eps \bxi_1 + \eps^2 \bxi_2 + O(\eps^3), \\
\barbu_\vp &= \bu_0 + \eps^2 \barbu_{\vp 2} + O(\eps^3), \\
\lambda &= \tfrac{1}{2} \|\bx\|^2 + \eps^2 \lambda_2 + O(\eps^3),
 \end{align}
 \end{subequations}
anticipating some $O(1)$ and $O(\eps)$ terms. 
Introducing this into \eqref{Xi*} gives
\beq
\bxi_{1t} + \lie_{\bu_0} \bxi_1 = \bu_1
\label{xi1}
\eeq
at $O(\eps)$. Here we introduce the notation
\beq
\lie_{\bu_0} \bxi_1 = \bu_0 \bcdot \bnabla \bxi_1 - \bxi_1 \bcdot \bnabla \bu_0
\eeq
for the Lie derivative (or commutator) of vector fields. We note that the solution $\bxi_1$ satisfies $\av{\bxi_1}=0$ and $\bnabla \bcdot \bxi_1 = 0$. For divergence-free vector fields like $\bu_0$ and $\bxi_1$, the Lie derivative takes the alternative form
\beq
\lie_{\bu_0} \bxi_1 = \bnabla \times ( \bxi_1 \times \bu_0),
\label{Luxi}
\eeq
making it plain it is also divergence free.

Averaging the $O(\eps^2)$ terms in \eqref{Xi*} yields
\beq
\barbu_{\vp 2} = \av{\bxi_1 \bcdot \bnabla \bu_1} + \tfrac{1}{2} \av{\bxi_1 \bcdot \bnabla \bnabla \bu_0 \bcdot \bxi_1} - (\partial_t + \lie_{\bu_0}) \av{\bxi_2},
\label{u*2}
\eeq
where we use that $\av{\bxi}_1 = 0$. The middle term on the right-hand side of \eqref{u*2} has the component expression
\beq
\av{\bxi_1 \bcdot \bnabla \bnabla \bu_0 \bcdot \bxi_1}^i = \av{\bxi_{1}^j \bxi_{1}^k} \partial_{jk} u_0^i,
\eeq
with summation over repeated indices implied. This term can be rewritten by noting that
\beq
\bxi_1 \bcdot \bnabla \bnabla \bu_0 \bcdot \bxi_1 = (\partial_t + \lie_{\bu_0}) (\bxi_1 \bcdot \bnabla \bxi_1) - \bu_1 \bcdot \bnabla \bxi_1 - \bxi_1 \bcdot \bnabla \bu_1.
\label{nablanablau0}
 \eeq
This identity can established in components starting with the first term on the right-hand side:
\begin{align}
\left( (\partial_t + \lie_{\bu_0}) (\bxi_1 \bcdot \bnabla \bxi_1) \right)^i &= (\partial_t + u_0^j \partial_j) (\xi_1^k \partial_k \xi_1^i) - \xi_1^k \partial_k \xi_1^j \partial_j u_0^i \nonumber \\
&= (\partial_t + u_0^j \partial_j) \xi_1^k \, \partial_k \xi_1^i + \xi_1^k \partial_k ((\partial_t + u_0^j \partial_j) \xi_1^i) \nonumber \\
& \quad - \xi_1^k \partial_j \xi^i \partial_k u_0^j - \xi_1^k \partial_k \xi_1^j \partial_j u_0^i \nonumber \\
&= u_1^k \partial_k \xi_1^i  + \xi_1^j \partial_j u_0^k \partial_k \xi_1^i + \xi_1^k \partial_k (u_1^i + \xi_1^j \partial_j u_0^i) \nonumber \\
& \quad - \xi_1^k \partial_j \xi^i \partial_k u_0^j - \xi_1^k \partial_k \xi_1^j \partial_j u_0^i \nonumber \\
&= u_1^k \partial_k \xi_1^i + \xi_1^k \partial_k u_1^i + \xi_1^j \xi_1^k \partial_{ik} u_0^i,
\end{align}
using \eqref{xi1} in the form $(\partial_t + u_0^j \partial_j) \xi_1^k = u_1^k + \xi_1^j \partial_j u_0^k$. Rearranging yields \eqref{nablanablau0}.

 Introducing \eqref{nablanablau0} into \eqref{u*2} gives
 \beq
\barbu_{\vp 2} = \tfrac{1}{2} \av{\lie_{\bxi_1} \bu_1} + (\partial_t + \lie_{\bu_0}) \left( \tfrac{1}{2} \av{\bxi_1 \bcdot \bnabla \bxi_1} - \av{\bxi_2} \right).
\label{u*22}
\eeq
Now, according to \eqref{lambdagrad}, $\av{\bxi_2}$ is a gradient, $\av{\bxi_2} = \nabla \lambda_2$, with $\lambda_2$ determined by the condition $\bnabla \bcdot \barbu_{\vp 2} = 0$. Since $\bnabla \bcdot \av{\lie_{\bxi_1} \bu_1}=0$, this implies
\beq
 \tfrac{1}{2} \av{\bxi_1 \bcdot \bnabla \bxi_1} - \av{\bxi_2} =  \tfrac{1}{2} \av{\bxi_1 \bcdot \bnabla \bxi_1} - \nabla \lambda_2 =  \tfrac{1}{2} \mathsf{P} \av{\bxi_1 \bcdot \bnabla \bxi_1},
\eeq
where $\mathsf{P}$ denotes the (Leray) projection on divergence-free vector fields.  This reduces \eqref{u*22} to
\beq
\barbu_{\vp 2} = \tfrac{1}{2} \av{\lie_{\bxi_1} \bu_1} + (\partial_t + \lie_{\bu_0}) \tfrac{1}{2}    \mathsf{P}  \av{\bxi_1 \bcdot \bnabla \bxi_1}
\label{u*23}
\eeq
and recovers the result of the `optimal transport' formulation of \cite{gilbert2018geometric}.

We can now compare the volume-preserving Lagrangian mean velocity $\barbu_\vp = \bu_0 + \eps^2 \barbu_{\vp 2}$ resulting form the polar factorization \eqref{polarfact}  with other Lagrangian mean velocities. The standard GLM Lagrangian mean velocity is $\barbu = \bu_0 + \eps^2 \barbu_{2}$ with
\beq
\barbu_{2} = \tfrac{1}{2} \av{\lie_{\bxi_1} \bu_1} + (\partial_t + \lie_{\bu_0})   \tfrac{1}{2}  \av{\bxi_1 \bcdot \bnabla \bxi_1}.
\label{GLM}
\eeq
It is obtained by setting $\av{\bxi_2}=0$ in \eqref{u*22}. 
Comparing with \eqref{u*23} shows that the  polar factorization construction  introduces the projection $\mathsf{P}$ in the last term. The glm construction of \citet{sowa-robe10}, reducing to the solenoidal Stokes drift of \citet{vanneste2022stokes} for $\bu_0=0$, is 
\beq
\barbu_2^{\mathrm{glm}} = \tfrac{1}{2} \av{\lie_{\bxi_1} \bu_1}. 
\label{glm}
\eeq

An important point is that the three Lagrangian mean velocities \eqref{u*23}, \eqref{GLM} and \eqref{glm} differ by terms that are derivatives along the leading order flow, $\partial_t + \lie_{\bu_0}$. This implies that the mean trajectories corresponding to the three mean velocities are $O(\eps^2)$ close for long times.  More precisely, we have that 
\beq
\barbphi_\vp(\ba,t) = \barbphi^{\mathrm{glm}}(\ba,t) + \eps^2 \bm{f}(\barbphi^{\mathrm{glm}}(\ba,t),t) + O(\eps^3),
\label{f}
\eeq
where $\barbphi^{\mathrm{glm}}$ is the flow map associated with $\bu_0 + \eps^2 \barbu^{\mathrm{glm}}$ and $\bm{f} =  \tfrac{1}{2}  \mathsf{P}  \av{\bxi_1 \bcdot \bnabla \bxi_1}$.

We verify \eqref{f} by taking the time derivative of the right-hand side to obtain
\begin{align}
& \barbu^\mathrm{glm}(\barbphi^{\mathrm{glm}}(\ba,t),t) + \eps^2 ( \partial_t \bm{f} + \barbu^\mathrm{glm} \bcdot \bnabla \bm{f})(\barbphi^{\mathrm{glm}}(\ba,t),t) + O(\eps^3) \nonumber \\
= \, &  \bu_0(\barbphi_\vp(\ba,t) - \eps^2 \bm{f}(\barbphi_\vp(\ba,t),t),t) \nonumber \\
& + \eps^2 \left(\tfrac{1}{2} \av{\lie_{\bxi_1} \bu_1} +  \partial_t \bm{f} + \bu_0 \bcdot \bnabla \bm{f} \right)(\barbphi_\vp(\ba,t),t) + O(\eps^3) \nonumber \\
= \, & \bu_0(\barbphi_\vp(\ba,t),t) + \eps^2 \left(\tfrac{1}{2} \av{\lie_{\bxi_1} \bu_1} +  (\partial_t  + \lie_{\bu_0}) \bm{f} \right) (\barbphi_\vp(\ba,t),t) + O(\eps^3) \nonumber \\
= \, & \barbu_\vp(\barbphi_\vp(\ba,t),t) + O(\eps^3).
\end{align}
This matches the time derivative $\partial_t \barbphi_\vp(\ba,t) = \barbu_\vp(\barbphi_\vp(\ba,t),t)$ of the left-hand side. 

A similar computation shows that
\beq
\barbphi(\ba,t) = \barbphi^{\mathrm{glm}}(\ba,t) + \eps^2 \bm{f}(\barbphi^{\mathrm{glm}}(\ba,t),t) + O(\eps^3),
\label{f2}
\eeq
with $\bm{f} =  \tfrac{1}{2}    \av{\bxi_1 \bcdot \bnabla \bxi_1}$.  

We conclude from the closeness of $\barbphi$, $\barbphi_\vp$ and $\barbphi^\mathrm{glm}$ shown by \eqref{f} and \eqref{f2} that they represent the mean motion equally well. 

We contrast this closeness  with the behaviour of the trajectories of a superficially appealing divergence-free alternative to $\barbu$, namely its projection $\mathsf{P} \barbu = \bu_0 + \eps^2 \mathsf{P} \barbu_2$ on divergence-free vector fields. Since
\beq
\mathsf{P} \barbu_2 = \tfrac{1}{2} \av{\lie_{\bxi_1} \bu_1} +   \mathsf{P}  (\partial_t + \lie_{\bu_0})  \tfrac{1}{2} \av{\bxi_1 \bcdot \bnabla \bxi_1} 
\eeq
and $\mathsf{P}$ and $\lie_{\bu_0}$ do not commute, the difference between $\mathsf{P} \barbu$ and $\barbu$ is not a derivative along the leading order flow. As a result, the trajectories of $\mathsf{P} \barbu$ do not remain close to $\barbphi$; they do not capture the mean motion and lead to the uncontrolled growth of the displacement field $\bxi$.

\section{Volume-preserving Lagrangian averaging with a Butterworth filter} \label{app:butter}

Volume-preserving Lagrangian averaging is not limited to the exponential filter of \S\ref{sec:exponential_mean}. It can be extended to a broad class of filters including the top-hat filter of Ref.\ \citep{kafiabad2023computing} or the 2nd-order Butterworth filter of Ref.\ \citep{minz2025efficient}. The key step is to identify the bare mean flow map $\langle \bphi\rangle = \barbphi$ and substitute the polar  decomposition $\bnabla \lambda \circ \barbphi_\vp$. 

To illustrate this, we apply the construction to the 2nd-order Butterworth filter,  an efficient low-pass filter in the class of sum-of-exponential filters considered by \citet{minz2025efficient}. 
For this filter, the mean flow map $\barbphi$ can be obtained by solving the system of equations
\begin{equation}
    \partial_t \begin{bmatrix} \tilde\bphi(\ba,t) \\ \barbphi(\ba,t) \end{bmatrix} = - \alpha A \begin{bmatrix} \tilde\bphi(\ba,t) \\ \barbphi(\ba,t) \end{bmatrix} + \alpha \begin{bmatrix} \bphi(\ba,t) \\ 0 \end{bmatrix}, 
    \label{eq:phiA}
\end{equation}
where $\tilde{\bphi}$ is an auxiliary field and the matrix
\begin{equation}\label{eq:multi_Matrix_A}
 A = \begin{bmatrix} \sqrt{2} - 1 & 2 - \sqrt{2} \\ -1 & 1 \end{bmatrix}
\end{equation}
is positive definite. Defining $\tilde \bXi = \bx + \tilde \bxi$ by  $\tilde \bphi =\tilde \bXi \circ \barbphi$, this can be rewritten as 
\begin{subequations} \label{eq:multi_xi}
\begin{align}
    \partial_t \tilde\bxi + \barbu \bcdot \bm\nabla \tilde\bxi = \tilde{\bu} - \barbu, \quad 
    & \textrm{and} \quad \partial_t \bxi + \barbu \bm\cdot \bm\nabla \bxi = \bu \circ (\id + \bxi) - \barbu,  \label{ppp} \\
\textrm{where} \quad     \tilde{\bu} = \alpha (\bxi - & (\sqrt{2} -1)  \tilde\bxi) \quad \textrm{and} \quad 
\barbu = \alpha \tilde\bxi.
\end{align}
\end{subequations}
The auxiliary velocity $\tilde \bu$ is defined by $\tilde \bu =  \partial_t \tilde \bphi  \circ \barbphi^{-1}$ (see \cite{minz2025efficient}).

We now obtain a volume-preserving version of \eqref{eq:multi_xi}.
We substitute $\barbphi = \bnabla \lambda \circ \bar{\bphi}_\vp$ into \eqref{eq:phiA} and compose  with $\barbphi_\vp^{-1}$ to find
\begin{equation}
        \begin{bmatrix} \tilde{\bu}_\vp (\bx,t) \\ \partial_t \bnabla \lambda(\bx,t) + (\bar{\bu}_\vp(\bx,t)\cdot \bnabla)\bnabla \lambda(\bx,t)  \end{bmatrix} = - \alpha A \begin{bmatrix} \tilde{\bXi}_\vp(\bx,t) \\ \bnabla\lambda(\bx,t) \end{bmatrix} + \alpha \begin{bmatrix} \bXi_\vp(\bx,t) \\ 0 \end{bmatrix},\label{eq:multi-velocity}
\end{equation}
where $\tilde{\bu}_\vp = \partial_t \tilde{\bphi} \circ \bar{\bphi}_\vp^{-1}$, $\tilde{\bXi}_\vp = \tilde{\bphi} \circ \barbphi_\vp^{-1}$ and $\bXi_\vp =  \bphi \circ\barbphi^{-1}_\vp$.
With $\tilde \bXi_\vp = \bx + \tilde \bxi_\vp$ and $\lambda = |\bx|^2/2 + \lambda'$, the first row in \eqref{eq:multi-velocity} becomes
\begin{align}\label{eq:auxfield}
    \tilde{\bu}_\vp = \alpha \left( 
     \bxi_\vp - (\sqrt{2} -1)\tilde{\bxi}_\vp - (2 -\sqrt{2})\bnabla\lambda'
    \right).
\end{align}
where $\tilde{\bXi}_\vp = \bx + \tilde{\bxi}_\vp$ and $\bXi_\vp = \bx +\bxi_\vp$. The second row of \eqref{eq:multi-velocity} is
\begin{equation}
    \partial_t \nabla  \lambda' +  (\bar{\bu}_\vp \cdot \nabla) (\nabla \lambda) = \alpha (\tilde{\bxi}_\vp - \bnabla \lambda'). \label{PDE: BW_proj} 
\end{equation}
Together with the divergence-free condition $\nabla \cdot \barbu_{\vp} = 0$, \eqref{eq:auxfield} and \eqref{PDE: BW_proj} determine $\tilde{\bu}_\vp$, $\barbu_{\vp}$ and $\lambda'$ assuming $\tilde \bxi_\vp$ and $\bxi_\vp$ are known. Evolution equations for  $\tilde \bxi_\vp$ and $\bxi_\vp$ are obtained by time differentiation of $\tilde \bphi = \tilde \bXi_\vp \circ \barphi_\vp$ and $\bphi = \bXi_\vp \circ \barphi_\vp$ and  are given by \eqref{ppp} with all fields decorated by $\dag$.

In two dimensions, we can take $\barbu_\vp =  \bnabla^\perp \psi$. The divergence and curl of \eqref{PDE: BW_proj} then give the system
\begin{subequations}
\begin{align}
\partial_t \nabla^2 \lambda' + \bnabla \bcdot \left( (\barbu_\vp \bcdot \bnabla) \bnabla \lambda' \right) &= \alpha (  \bnabla \bcdot \tilde \bxi_\vp - \nabla^2 \lambda'),  \\
\nabla^2 \bar \psi_\vp + \bnabla \times \left((\barbu_\vp \bcdot \bnabla) \bnabla \lambda' \right) &= \alpha \bnabla \times \tilde \bxi_\vp. 
\end{align}
\end{subequations}
analogous to \eqref{lampsi}.

\bibliography{ref}

\end{document}